\documentclass[aps,twocolumn]{revtex4-1}
\usepackage{amssymb}
\usepackage{amsmath,bm}
\usepackage{graphicx,color}
\usepackage{bbm}
\usepackage{multirow}
\usepackage{hyperref}
\usepackage[justification=justified,width=\linewidth]{caption}
\usepackage{subcaption}
\begin{document}
\title{Phase separation and aging dynamics of binary liquid in porous media}
\author {Rounak Bhattacharyya}
\author {Bhaskar Sen Gupta}
\email{bhaskar.sengupta@vit.ac.in}
\affiliation{Department of Physics, School of Advanced Sciences, Vellore Institute of Technology, Vellore, Tamil Nadu - 632014, India}
\date{\today}

\begin{abstract} 
We employ the state-of-the-art molecular dynamics simulations to study the kinetics of phase separation and aging phenomena of segregating binary fluid mixtures imbibed in porous materials. Different random porous structures are considered to understand the effect of pore morphology on coarsening dynamics. We find the effect of complex geometrical confinement resulting in the dramatic slowing down in the phase separation dynamics. The domain growth follows the power law with an exponent dependent on the porous host structure. After the transient period, a crossover to a slower domain growth is observed when the domain size becomes comparable to the pore size. Due to the geometric confinement, the correlation function and structure factor modify to a non-Porod behavior and violate the superuniversality hypothesis. The role of porous host structure on the nonequilibrium aging dynamics is studied qualitatively by computing the two-time order-parameter autocorrelation function. This quantity exhibits scaling laws with respect to the ratio of the domain length at the observation time and the age of the system. We find the scaling laws hold good for such confined segregating fluid mixtures.
\end{abstract}
\maketitle
The phase separation of liquids imbibed in porous material has enjoyed increased theoretical and experimental 
attention because of their great scientific interest \cite{Brochard,Gennes,Maher,Goh} and industrial applications \cite{Kanamori} especially in the oil recovery process \cite{Morrow}. The 
geometric confinement and the porous host structure play an important role in the demixing behavior. The effect of 
pore size on the coarsening process is well studied and can be classified into two broad categories depending on the 
ratio of two major length scales involved in the system: the thermal correlation length $\xi$ of the liquid to the average pore diameter $d_p$. If $\xi \le d_p$, the pores act as a quenched random field, analogous to magnets with random impurities \cite{Brochard,Gennes}. Therefore, the system is well described by the random field Ising model. In the case where $\xi > d_p$, the pore size, and topology are expected to dictate the demixing process.

Systems with $\xi > d_p$ appear to be relevant for industrial application purposes. Attempts were made to understand
the role of the pore size in the coarsening process using cylindrical pore geometry \cite{Liu,Liu1,Tanaka}. However, the effect of the complex topology of the pore structure is still in its infancy. The reason can be attributed to the experimental limitation of probing the real space geometry using scattering experiments. Also, the theoretical and numerical studies are impeded by the complex topology of the material. Only fewer experiments were carried out to investigate the impact of complex geometry on phase separation \cite{ Kanamori,Iglauer,Pak}. Computer simulations employed using the standard Molecular Dynamics algorithm \cite{Strickland}, lattice Boltzmann \cite{Grunau} and phase-field method \cite{Chakrabarti} revealed rather an early stage of phase separation only in 2d porous material. It is expected that the geometrical randomness and the absence of translational invariance on long-length scales will play an important role in altering the phase separation kinetics of fluid mixtures imbibed in porous materials. However, this topic is less explored and poorly understood. In particular, the long-time growth dynamics during the coarsening and the aging properties of these nonequilibrium processes are completely unexplored. 

 The main aim of this work is to investigate the effect of the complex pore topology of the host material on the phase separation kinetics and aging phenomena of the fluid. We address these topics on phase separation by employing extensive computer simulations of immiscible symmetric binary fluid mixture embedded in random porous materials in 3d. The materials with a wide range of pore structures are formed analogously to vycor glasses.
 
 To study coarsening dynamics, we resort to molecular dynamics (MD) simulation in the NVT ensemble. For the binary liquid, a $50:50$ mixture of A and B particles is considered at high density $\rho= N/V=1$, where N and V represent the number of particles and volume of the system respectively. The two species interact via Lennard-Jones (LJ) potential
 $U_{\alpha\beta}(r) = 4\epsilon_{\alpha\beta}\Big[\Big(\frac{\sigma_{\alpha\beta}}{r}\Big)^{12} - \Big(\frac{\sigma_{\alpha\beta}}{r}\Big)^{6} \Big] $ where $r=|\vec{r_i}-\vec{r_j}|$ and $\alpha, \beta \in  \rm{A, B}$. To ensure energetically favorable phase separation, the parameters in the LJ potential are chosen as follows: $\sigma_{AA} = \sigma_{BB} = \sigma_{AB} = 1.0$ and $\epsilon_{AA} = \epsilon_{BB} = 1.0, \epsilon_{AB} = 0.5$. The choice of our interaction strength corresponds to the critical temperature $T_c=1.42$, outlying the possible liquid-solid and gas-liquid transition point \cite{Das}. The temperature is measured in units of $\epsilon/k_B$, where $k_B$ is Boltzmann’s constant. Length and time are measured in units of $\sigma$ and $(m\sigma^2/\epsilon)^{1/2}$ respectively. For simplicity, we set the mass $m_0$ of A and B particles and $k_B$ equal to unity. For the sake of computational efficiency, the interaction potential is truncated to zero at $r_c=2.5\sigma$. Periodic boundary condition is applied in all three directions.
 
 The following method is adopted to form the porous media. We begin our simulation by preparing a well equilibrated homogeneous mixture of $N=262144$ ($64^3$) particles at high temperature $T=10.0$ followed by a quench to $T=0.77 T_c$. The system is then allowed to phase separate for a time period $\tau$ whereby interconnected domain structures of the same species form. At this juncture, the domains formed by one of the species are considered as porous host structure and the particles are relabeled as $P$ type. The $P$ type particles are kept frozen throughout the rest of the simulation. The particles of the other species are then randomly relabeled as $A$ and $B$ type keeping their number ratio $50:50$. The interaction potential for the $P$ particles is truncated at $r_c=2^{\frac{1}{6}}\sigma$ to exclude any attractive force. The sample is then heated up again to $T = 10.0$ for equilibration to annihilate any memory effect. Finally, the system is quenched to $T=0.77 T_c$ at $t=0$ and is allowed to evolve to the thermodynamically favored state until the phase separation is achieved. Temperature is controlled by the Nose-Hoover thermostat (NHT), which preserves the hydrodynamic effect. 
 
\begin{figure}
 	\centering
 	\begin{subfigure}[h!]{0.5\columnwidth}
 		\includegraphics[width=\columnwidth]{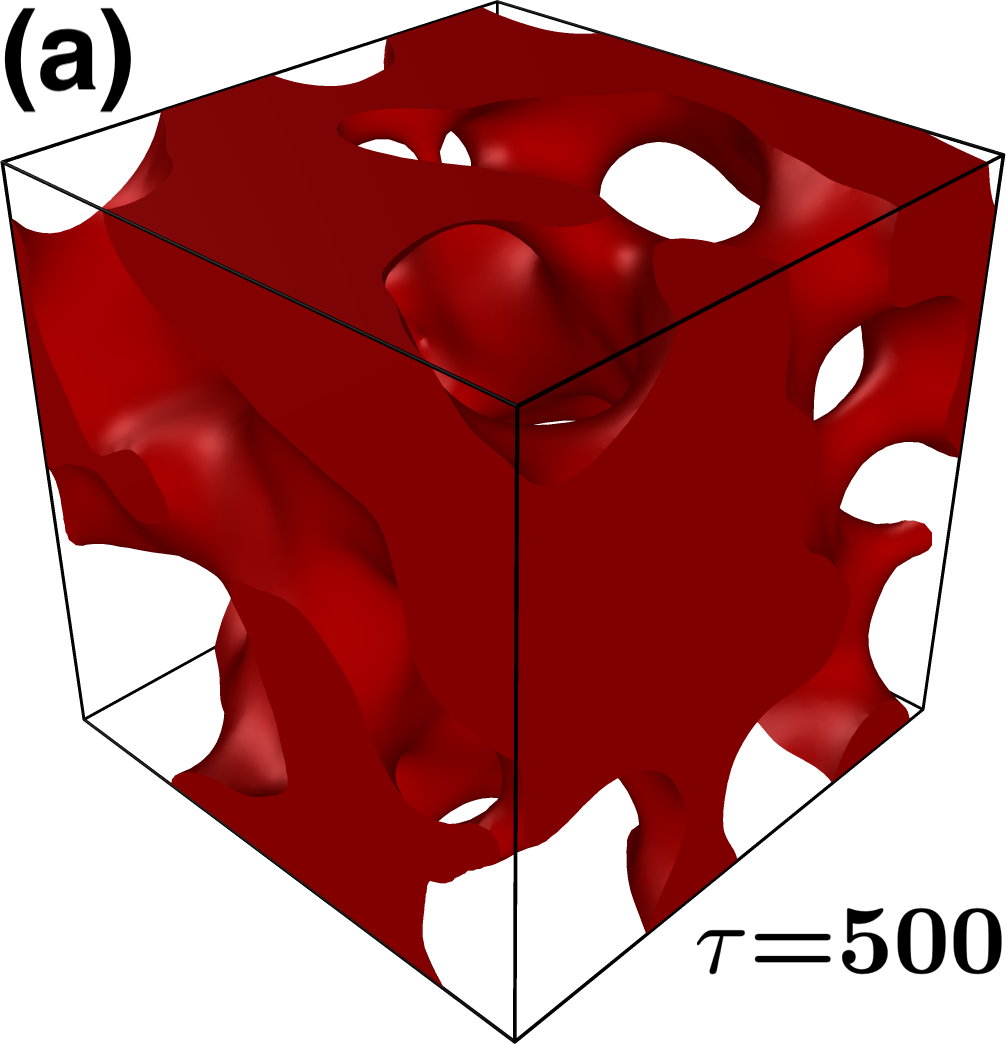}
 	\end{subfigure}%
 	\begin{subfigure}[h!]{.5\columnwidth}
 		\centering
 		\includegraphics[width=\columnwidth]{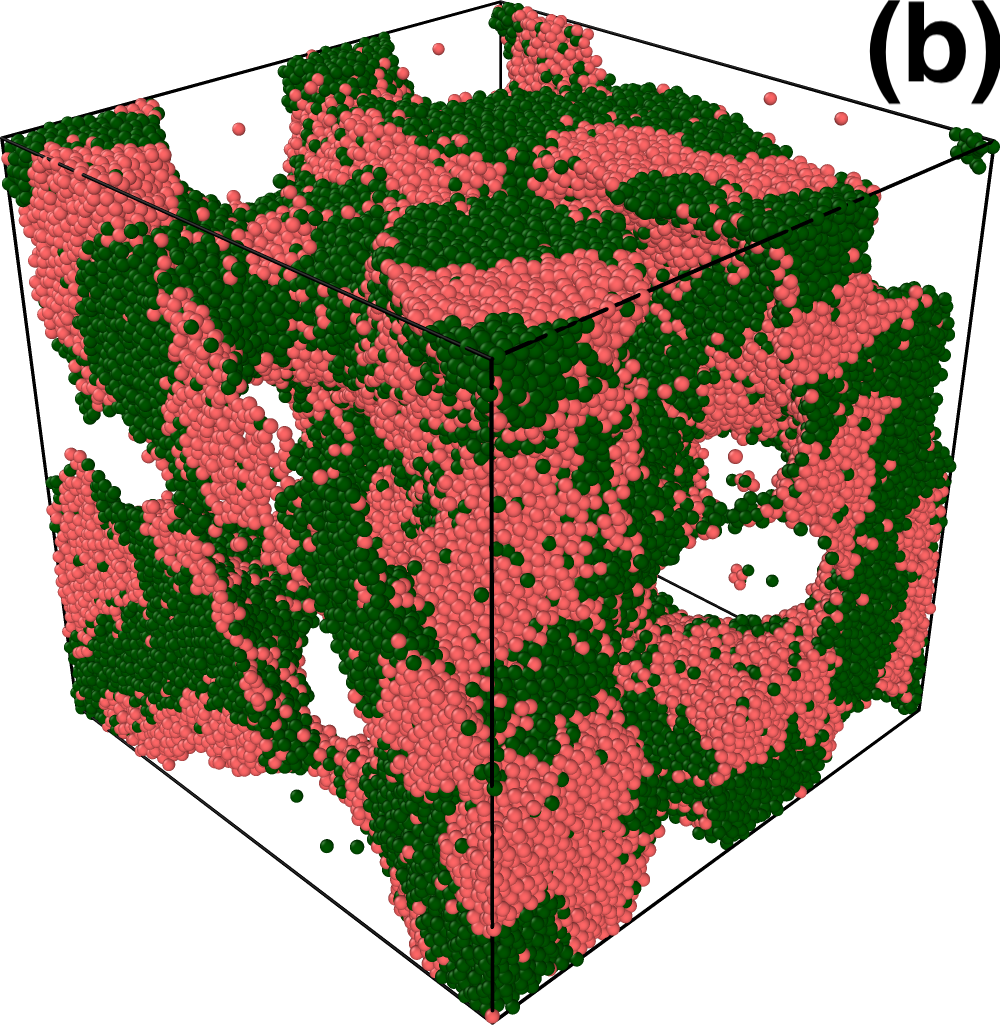}
 	\end{subfigure}\\
  	\begin{subfigure}[h!]{0.5\columnwidth}
 	\includegraphics[width=\columnwidth]{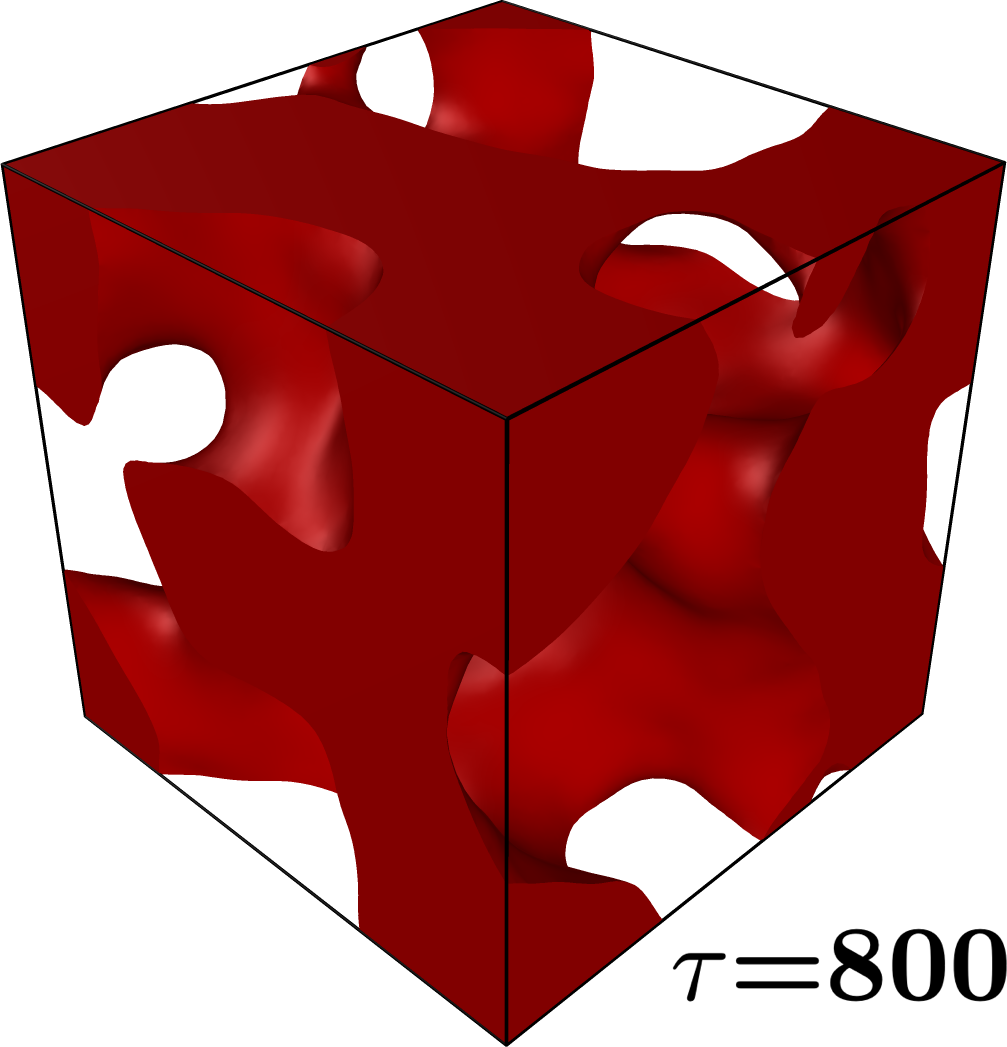}
 \end{subfigure}%
 \begin{subfigure}[h!]{.5\columnwidth}
 	\centering
 	\includegraphics[width=\columnwidth]{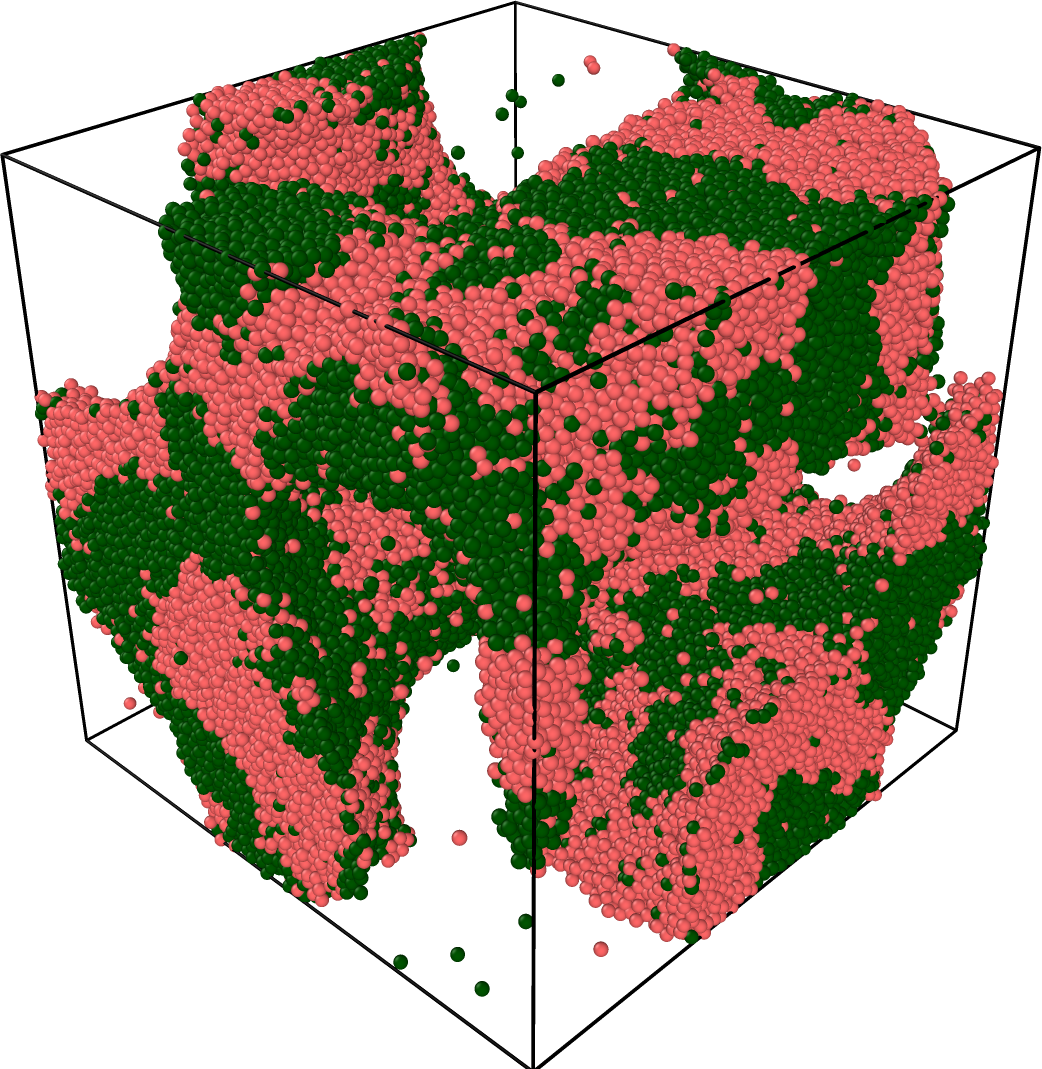}
 \end{subfigure}\\
 	\begin{subfigure}[h!]{0.5\columnwidth}
	\includegraphics[width=\columnwidth]{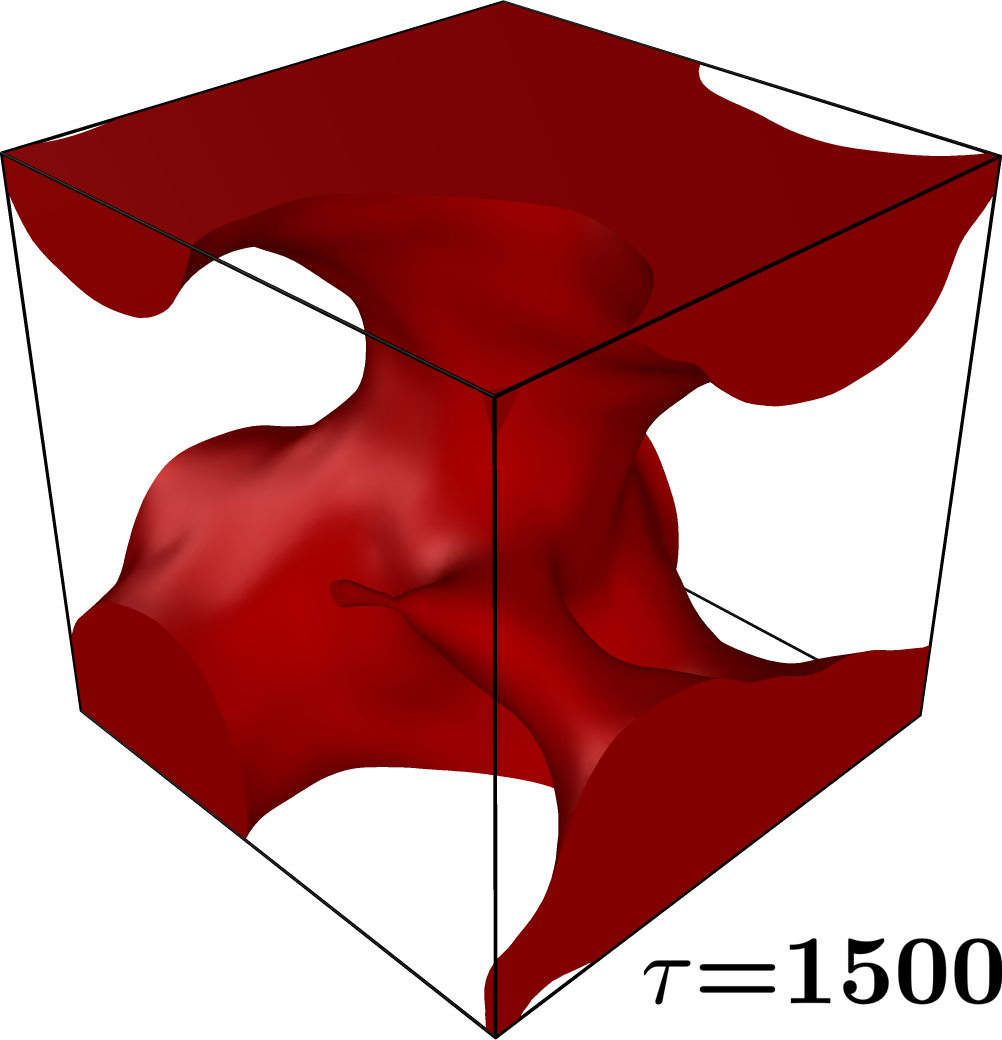}
\end{subfigure}%
\begin{subfigure}[h!]{.5\columnwidth}
	\centering
	\includegraphics[width=\columnwidth]{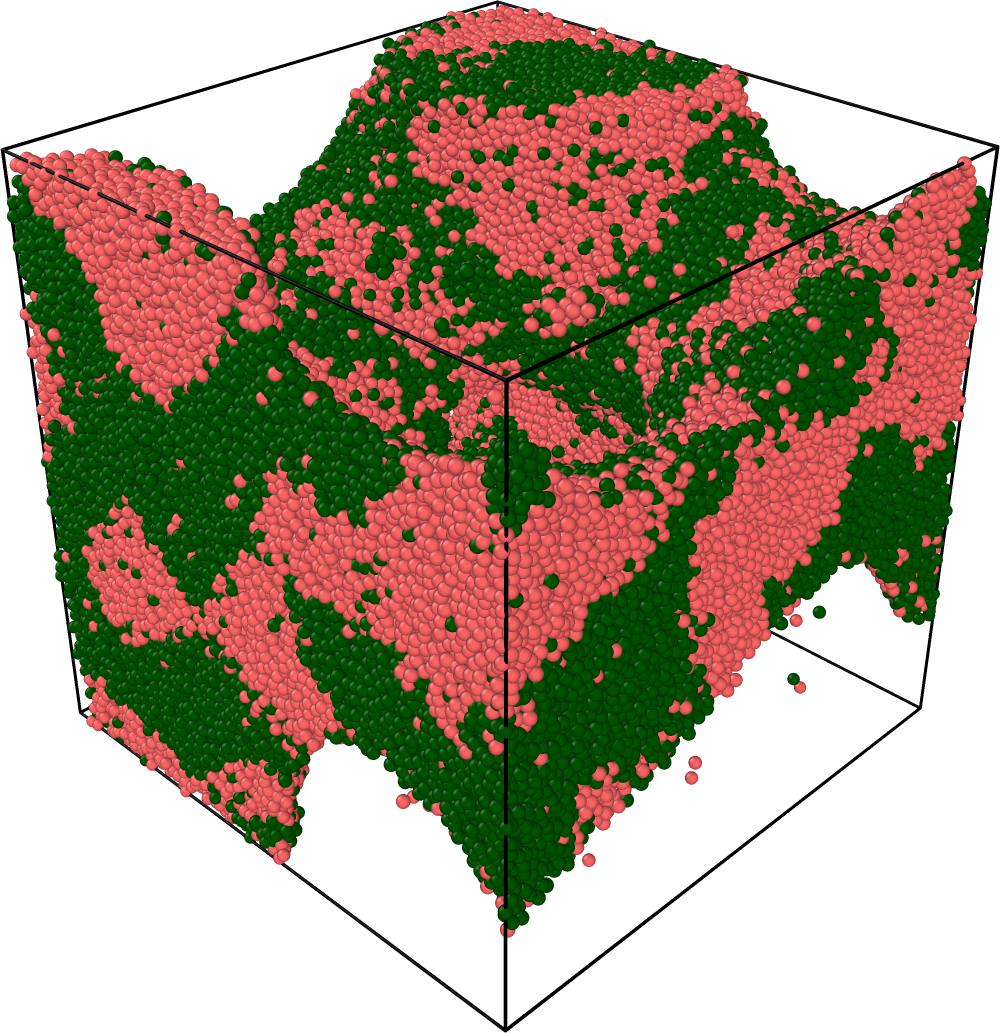}
\end{subfigure}
 	\caption{(Color online) (a) Three different porous host structures used in our simulations. (b) Typical snapshots of the  phase-separating binary liquid system at time $t=3000$. The $A$ and  $B$ particles are marked as black and orange respectively.}
 	\label{fig1-snap}
 \end{figure}

 As follows from the simulation protocol, the time period $\tau$ is directly related to the average pore size which translates to the mean size of the interconnected domains of $P$ type particles. In our present study, we consider three different porous host structures which correspond to $\tau$=500, 800, and 1500 as shown in Fig~\ref{fig1-snap}(a). From the figure, it is evident that the average pore size increases with an increase in $\tau$. The effect of porous media on the phase separation dynamics is depicted in Fig~\ref{fig1-snap}(b). For all the cases bicontinuous $A$-rich and $B$-rich domains are observed. It is apparent from the snapshots that the same species cluster sizes get larger with an increase in the average pore size at a given time. This indicates the kinetics of ordering are strongly dependent on the pore size.

 To characterize the domain morphology and study the domain growth of the segregating mixture imbibed into these  
porous media we introduce the two-point equal time correlation function $C_{\psi\psi}(\vec{r},t)$ given by 
 \begin{equation}\label{Correlation_function}
 C_{\psi\psi}(\vec{r},t) = \langle\psi(0,t)\psi(\vec{r},t)\rangle /\langle\psi(0,t)\rangle^2
 \end{equation} 
 where the order parameter $\psi(\vec{r},t)$ is obtained as follows: we compute the local density difference $\delta\rho = \rho_A-\rho_B$ between the two species A and B, calculated over a box of size $(2\sigma)^3$
   located at $\vec{r}$. The $\psi(\vec{r},t)$ is assigned a value $+1$ when $\delta\rho>0$, and $-1$ otherwise. The angular brackets stand for statistical averaging. The structure factor $S(\vec{k},t)$ is computed by taking the 
   Fourier transform of the correlation function given by $S(\vec{k},t) = \int d\vec{r} \hspace{0.1cm} exp(i\vec{k}.\vec{r}) \hspace{0.1cm} C_{\psi\psi}(\vec{r},t)$. Finally for the isotropic system, spherically 
   averaged $C_{\psi\psi}(r,t)$, and $S(k,t)$ are calculated.
   
    \begin{figure}[h!]
   	\centering
   	\includegraphics[width=\columnwidth]{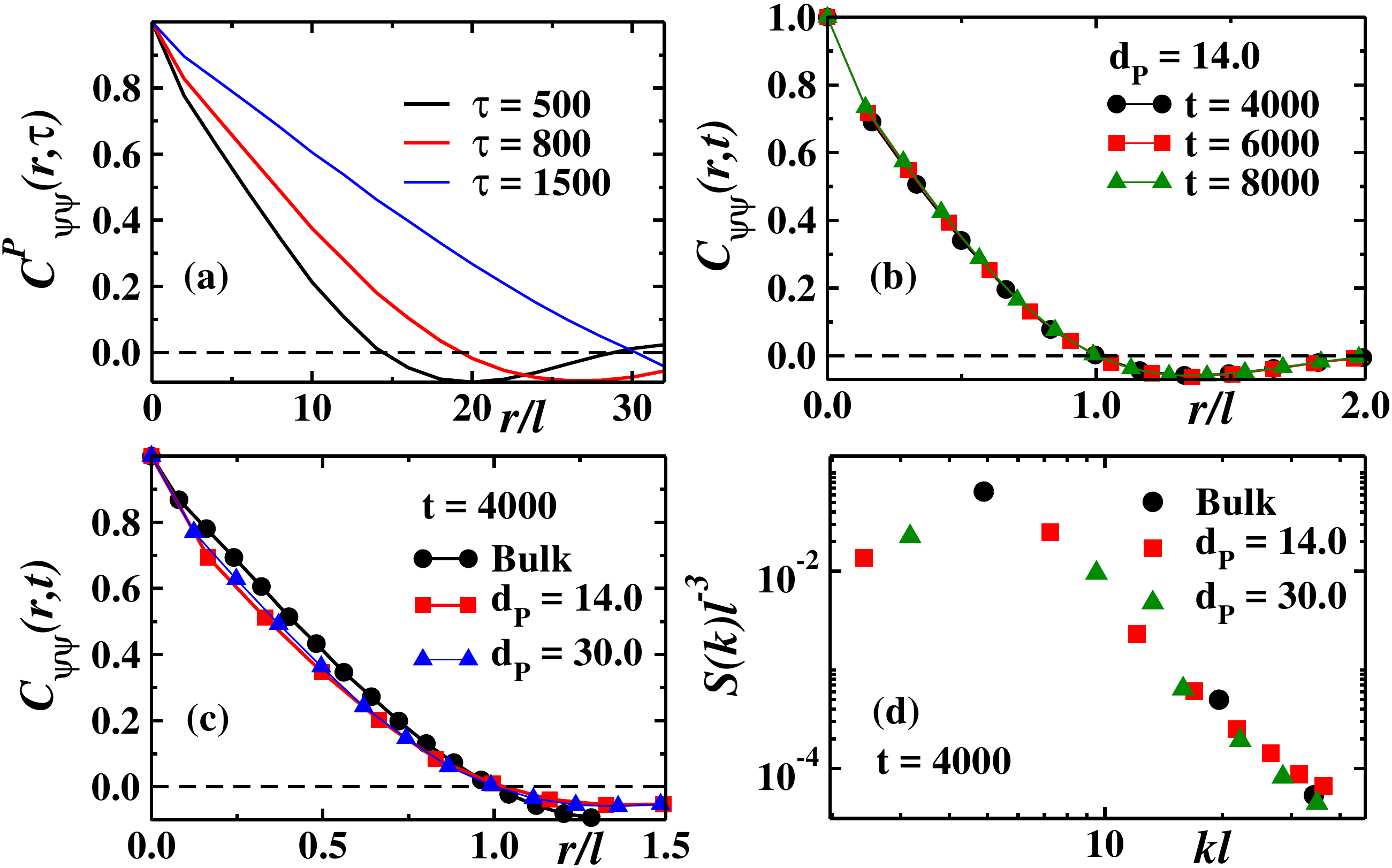}
   	\caption{(a) Scaled correlation function $C_{\psi\psi}^{P}(r,t)$  vs $r/\ell(t)$ for the $P$ type particles constituting three different porous host structures. (b) The scaling plot of $C_{\psi\psi}(r,t)$ vs $r/\ell(t)$ for the liquid mixture inside the porous material with $d_p=14.0$. (c) The same scaling plot in Fig. (b) for the liquid in bulk and confined in porous media with $d_p=14.0$ and $30.0$. (d) The scaled structure factor $S(k)\ell^{-3}$ vs $\ell k$ is plotted for the bulk liquid and inside the porous material with $d_p$ = 14.0 and 30.0}
   	\label{cor}
   \end{figure}
   
 To gain a qualitative understanding of the average domain size we resort to the correlation function given by Eq. \ref{Correlation_function}. We first measure the average pore size of different host materials shown in Fig~\ref{fig1-snap}(a) which is quantified as the average domain size of the $P$ particles. For that we plot in Fig~\ref{cor}a, the correlation function $C^P_{\psi\psi}(r,\tau)$ for the $P$ particles vs $r/ \ell(\tau)$ for $\tau=500, 800, \mathrm{and}~1500$. Here $\ell(\tau)$ is the average domain size, measured through the first zero of $C^P_{\psi\psi}(r,\tau)$. From the figure we obtain the average domain size ($d_p$) as 14.0 ($\tau=500$), 19.0 ($\tau=800$) and 30.0 ($\tau=1500$).
 
  In Fig~\ref{cor}b we show the scaling plot of the correlation function for the binary fluid $C_{\psi\psi}(r,t)$ vs  $r/ \ell(t)$ confined in one of the pore structure $d_p=14.0$. An excellent data collapse is observed for different times. A similar scaling behavior is observed for other pore structures ($d_p=$ 19.0 and 30.0) also (not shown). This suggests that even being imbibed in the porous media, the segregating system belongs to the same dynamical universality class \cite{Binder3}. However, the correlation functions corresponding to different pore morphology do not overlap with each other when plotted at a fixed time $t$. This is shown in Fig.~\ref{cor}c. For the bulk system, the $C_{\psi\psi}(r,t)$ exhibits a linear decay at short length scale following the Porod law \cite{Porod}, which results from scattering off sharp interfaces. However, inside the porous structure, a nonlinear or cusp nature is observed, which can be attributed to the scattering from the modified domain boundaries \cite{Gaurav,Gaurav1}. Therefore, the presence of porous medium results in breaking down the Porod law in the correlation function \cite{Shaista}. 
 
 An important question in this regard is the validity of the so-called superuniversality (SU), i.e, whether the effect of porous media is limited to the domain growth law only, or there is also an explicit pore structure dependence. We, therefore, verify whether the spatial autocorrelation function $C_{\psi\psi}(r,t)$ corresponding to different pore morphology scales to a master curve when the length is rescaled with respect to the domain size $\ell(t)$ \cite{Lai,Corberi}.  From Fig.~\ref{cor}(c) it is clear that the correlation functions do not scale onto a master curve. Therefore, the systems into consideration do not belong to the SU class. The violation of SU hypothesis is further confirmed from the structure factor as shown in Fig~\ref{cor}(d) where we show the scaled $S(k,t) \ell^{-3}$ vs $\ell k$ in log-log scale. We observe the decaying part of the tail of $S(k,t)$ for the liquid inside the porous system deviates from the Porod law behavior $S(k,t) \sim k^{-4}$ seen in the bulk \cite{Binder3}. 
 
 \begin{figure}[h!]
 	\centering
 	\includegraphics[width=\columnwidth]{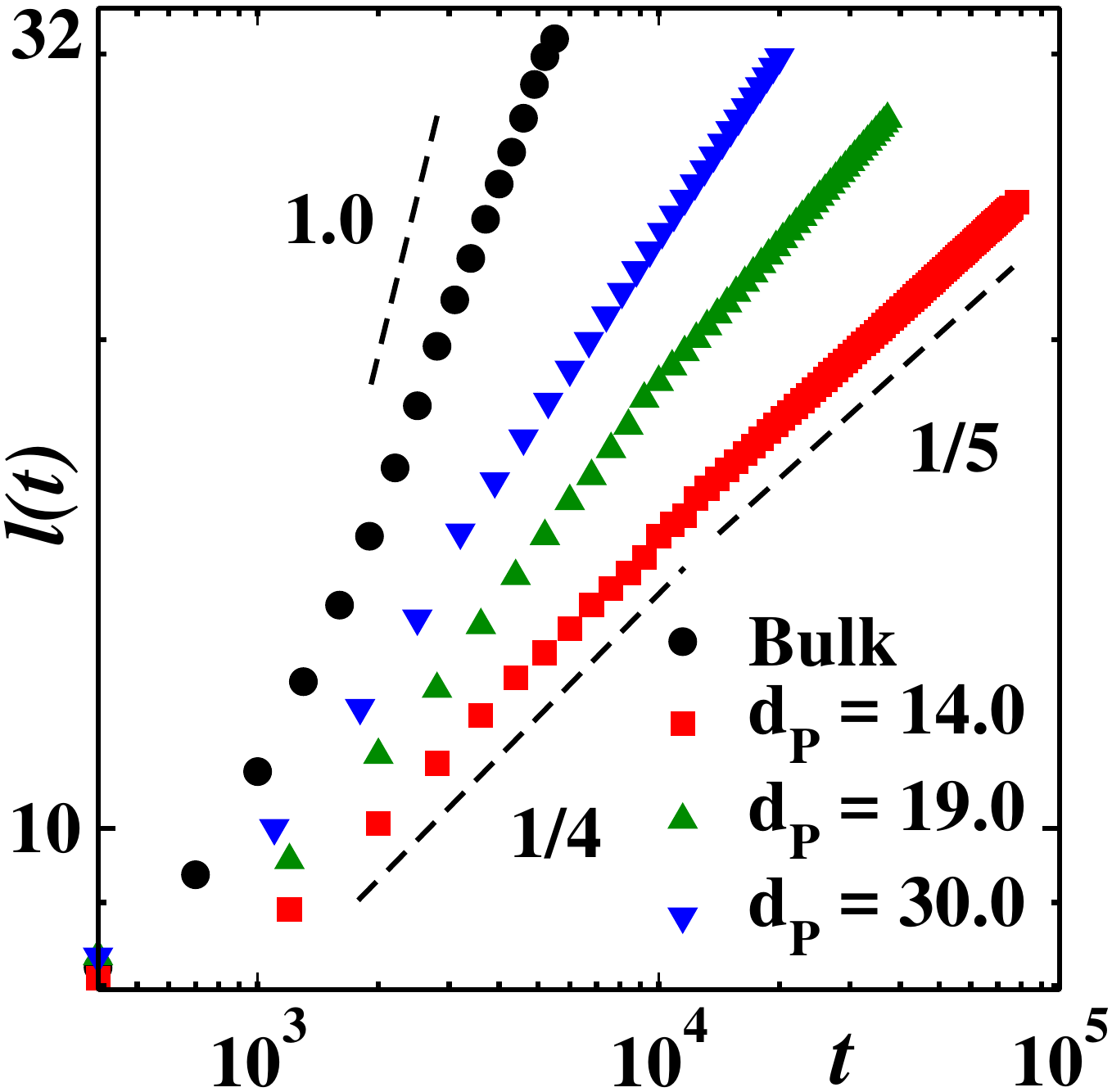}
 	\caption{The time evolution of the average domain size $\ell(t)$ for the bulk system and the mixture inside different porous structures characterized by the average pore diameter $d_p$  in the log-log scale. The dashed lines represent the guideline for the slope.}
 	\label{lscale}
 \end{figure}

 To understand the evolution quantitative, we focus on the time dependence of the domain size $\ell(t)$. In Fig.~\ref{lscale} we show the $\ell(t)$ for different pore sizes ($d_p=14.0, 19.0, \mathrm{and}~30.0$) with time. Our simulation is able to access the viscous hydrodynamic regime after an initial transient period. For the bulk system, the average domain size is expected to grow with time as $\ell(t) \sim t^\alpha$ with the exponent $\alpha=1$ in the viscous hydrodynamic regime \cite{Siggia,Furukawa}. The slight deviation from this can be attributed to the nonzero off-sets at the crossovers which can be subtracted from $\ell(t)$ to recover the proper linear behavior \cite{skd}. Inside the porous media, the coarsening dynamics of the segregating liquid slows down and the total time taken for the system to be phase-separated increases significantly with decreasing pore size. This is consistent with the observations in Fig.~\ref{fig1-snap}. 
 
 Careful observation reveals two different growth regimes in Fig.~\ref{lscale}. At the early stage, the average domain size is smaller than the pore size $d_p$ and the growth is relatively faster. Note that, in this regime also spatial confinement plays an important role and the growth is slower compared to the bulk. As the average domain size becomes comparable with $d_p$, the growth slows down further. This crossover happens at an earlier time for the systems with smaller $d_p$. The slowing down of the coarsening dynamics inside the porous medium compared to the bulk can be comprehended as follows. In the bulk, the driving force of coalescence depends on the difference in curvature of the neighboring domain interfaces. In the presence of the porous material, the domain interface is completely determined by the porous structure. Therefore the confining geometry shows down the phase separation dynamics significantly. We find the diffusive coalition process continuing steadily during our simulation time without any pause. The slowest growth for the system under consideration ($d_p=14.0$) corresponds to $\alpha=1/5$.
 
 \begin{figure}[]
	\centering
	\includegraphics[width=\columnwidth]{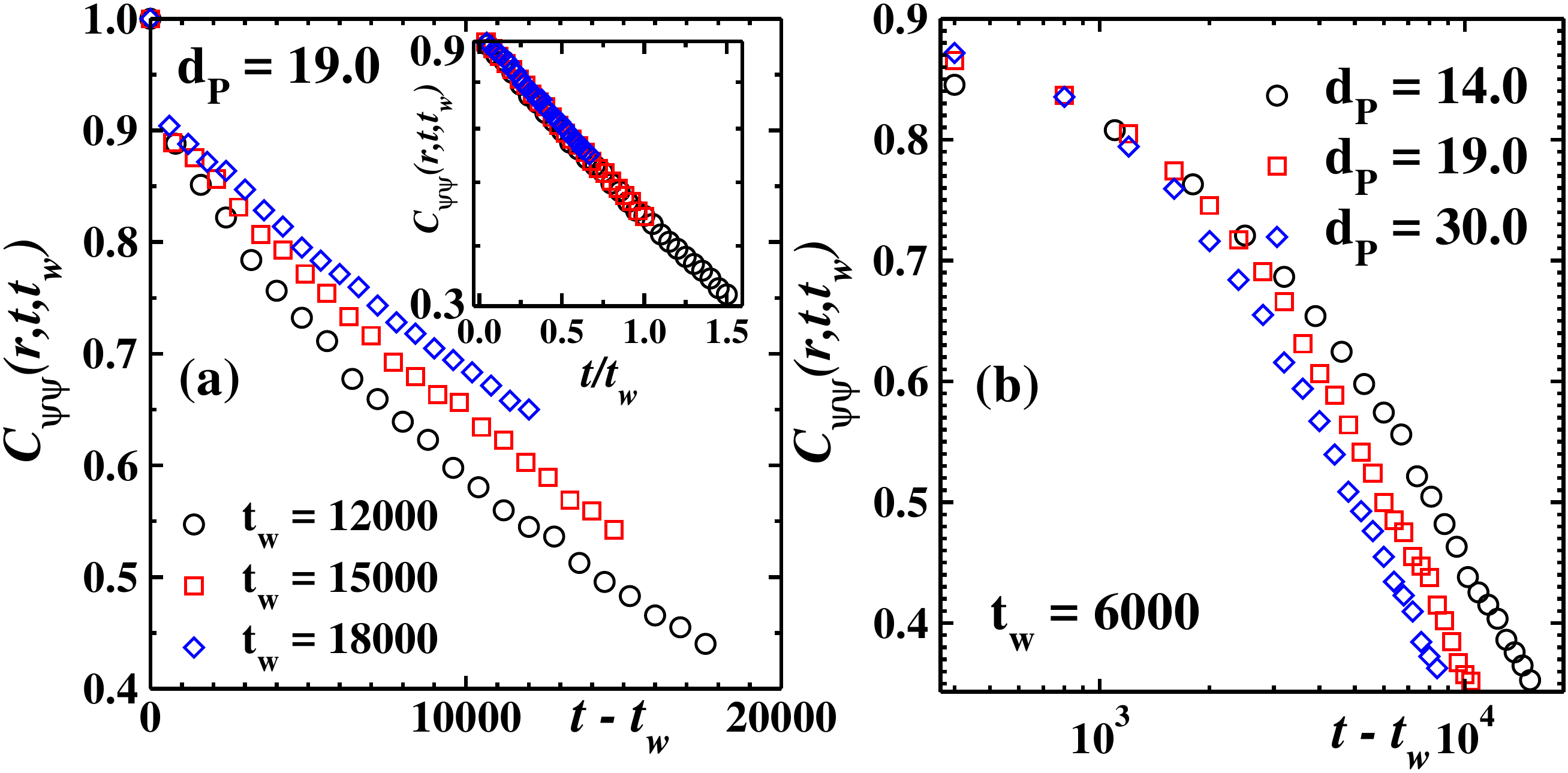}
	\caption{(a) The two-point order parameter autocorrelation function $C_{\psi\psi}(r,t,t_w)$ vs $t-t_w$ plot for the liquid mixture inside the porous structure with $d_p=19.0$ at three different waiting time $t_w=$ 12000, 15000, and 18000. In the inset, we show the same data after rescaling the time with $t_w$. (b) The plot of $C_{\psi\psi}(r,t,t_w)$ vs $t-t_w$ for the binary  system inside different pore structures at $t_w=6000$.}
	\label{aging}
\end{figure}

 One important topic related to the phase separation process is the aging phenomenon. Aging means the absence of time translation invariance and the older system relaxes slowly. This subject is well studied for the bulk \cite{sahmed} but remained completely unexplored in the porous media. For the aging-related studies during the non-equilibrium coarsening dynamics we resort to the so-called two-time order-parameter correlation function $C_{\psi\psi}(r,t,t_w)$ as follows:
\begin{equation}\label{autoCorrelation_function}
C_{\psi\psi}(r,t,t_w) = \langle\psi(\vec{r},t)\psi(\vec{r},t_w)\rangle - \langle\psi(\vec{r},t)\rangle\langle\psi(\vec{r},t_w)\rangle   
\end{equation}
where $t$ is the observation time and $t_w$ is the waiting time or the age of the system after the quench. For our present study, we always focus on the longer time regime after the crossover where the effect of spatial confinement is apparent (as reflected in Fig.~\ref{lscale}), and the $t_w$'s are chosen accordingly. In Fig.~\ref{aging}(a) we show the variation of $C_{\psi\psi}(r,t,t_w)$ with $t - t_w$ for the porous host structure with $d_p=19.0$. Clearly, the correlation curves corresponding to different $t_w$ do not overlap, demonstrating the violation of time translation invariance.  Following Fisher and Huse \cite{Fisher1} we attempt to scale the abscissa as $t/t_w$ and a nice data collapse is obtained as shown in the inset of Fig.~\ref{aging}(a). We repeated the same exercise for the systems confined into other pore structures $d_p= 14.0~ \mathrm{and}~30.0 $ (not shown) and observed the same behavior. Therefore, the Fisher and Huse scaling law remains vindicated for the segregating systems imbibed in the porous media.  To gain further insight in the aging dynamics the $C_{\psi\psi}(r,t,t_w)$ is plotted in Fig.~\ref{aging}(b) at a fixed $t_w=3000$ for different pore morphologies. The slowing down of the coarsening dynamics is evident with decreasing average pore size $d_p$ of the material and is consistent with the observation in Fig.~\ref{lscale}.
 \begin{figure}[h!]
	\centering
	\includegraphics[width=\columnwidth]{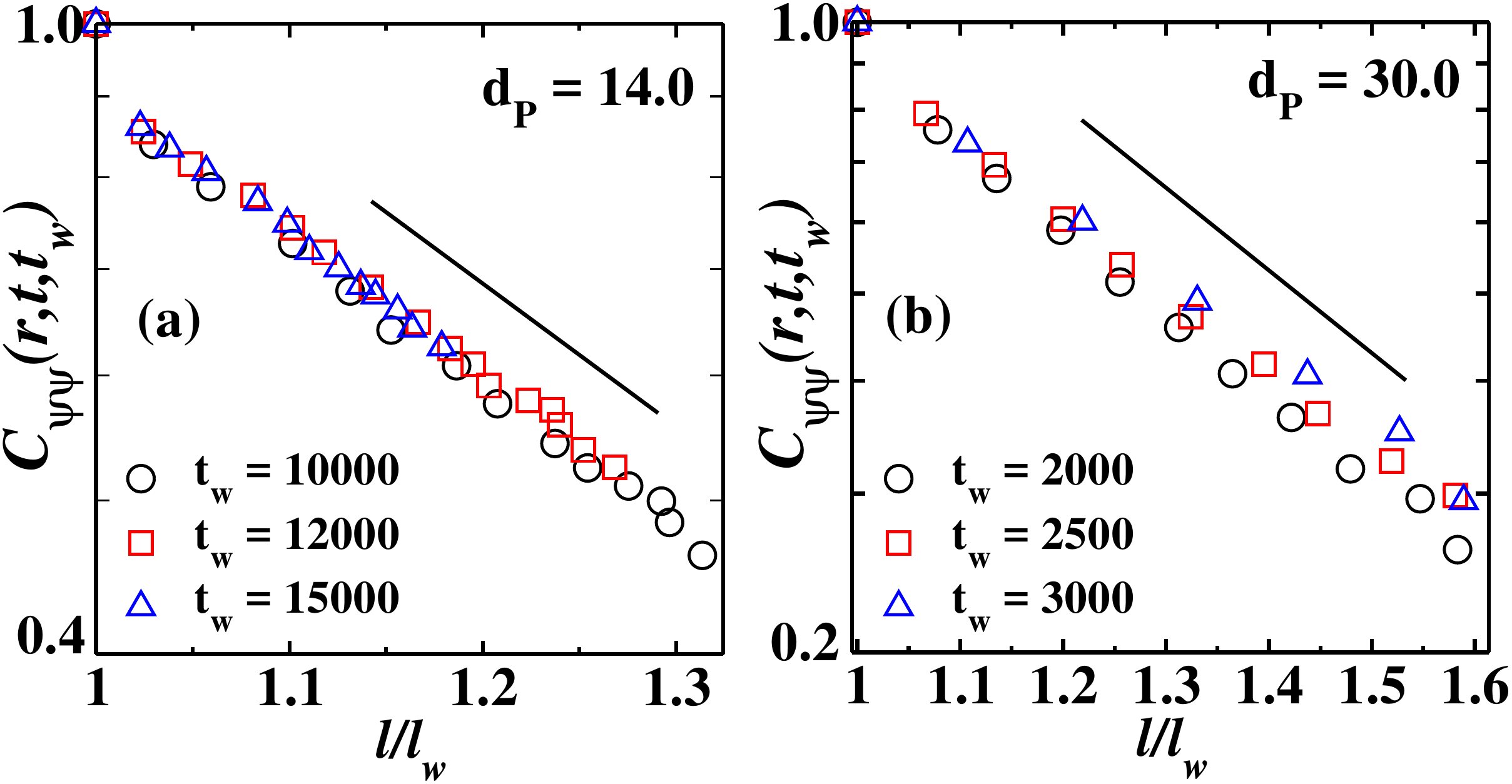}
	\caption{(a) The scaling of the correlation function $C_{\psi\psi}(r,t,t_w)$ plotted as a function of $l/l_w$ for the liquid system with porous host $d_p=14.0$ for different values of $t_w$ in the semi-log scale. (b) Same result as in (a) for the pore structure $d_p=30.0$.}
	\label{agingScale}
\end{figure}

In Fig.~\ref{agingScale} we show the $C_{\psi\psi}(r,t,t_w)$ as a function of $l/l_w$ for the pore structures with $d_p = 14.0~ \mathrm{and}~30.0$. An excellent data collapse is found for both the systems. To tally the nature of the master curve with the previously found exponential decay in the bulk system \cite{sahmed}, we plot the data on the semi-log scale. The data set appears to be linear confirming the exponential nature. Therefore, we find the scaling law in the hydrodynamic regime is very generic, insensitive to the pore topology. The exponential decay of the correlation function is attributed to the advective hydrodynamic flows under the hydrodynamic effect that causes large displacement of particles \cite{sahmed}.

In summary, we have studied the phase separation of segregating binary fluid imbibed in porous media using extensive MD simulations. Three different random porous structures were considered to understand the effect of pore morphology on coarsening dynamics. We observe a dramatic slowing down in the phase separation dynamics with decreasing average pore size. The spatial correlation function scaled with domain size showed pore size dependence and non-Porod behavior. Similar behavior is manifested be the decaying part of the tail of the structure factor. This can be attributed to the modification of domain boundaries due to geometrical confinement.  After an initial period of domain growth, when the domain size becomes comparable with $d_p$, a crossover to a slow-growth regime is observed.

We have also investigated the nonequilibrium aging dynamics in terms of the two-time order-parameter autocorrelation function $C_{\psi\psi}(r, t , t_w)$. We demonstrated that $C_{\psi\psi}(r,t,t_w)$ follows the scaling law with respect to $t/t_w$ proposed by Fisher and Huse in the disordered system. But the aging process slows down significantly as reflected in the decay of $C_{\psi\psi}(r,t,t_w)$ with time. We have also examined the scaling law with respect to $\ell/\ell_w$ and obtained an excellent data collapse for all the cases in hand. The scaling function showed an exponential decay in the hydrodynamic regime.

\noindent{\it Acknowledgment.---} B.S.G. acknowledges Science and Engineering Research Board (SERB), Dept. of Science and Technology (DST), Govt. of India (no. SRG/2019/001923) for financial support.


\begin{thebibliography}{99}
 
 \bibitem{Brochard}
 F. Brochard and P. G. de Gennes, J. Phys. Lett. (Paris) \textbf{44}, 785 (1983).
 \bibitem{Gennes}
 P. G. de Gennes, J. Phys. Chem. \textbf{88}, 6469 (1984).
 \bibitem{Maher}
 J. V. Maher, W. I. Goldburg, D. W. Pohl, and M. Lanz, Phys. Rev. Lett. \textbf{53}, 60 (1984).
 \bibitem{Goh}
 M. C. Goh, W. I. Goldburg, and C. M. Knobler, Phys. Rev. Lett. \textbf{58}, 1008 (1987).
 \bibitem{Kanamori}
 K. Kanamori, K. Nakanishi, and T. Hanada, Soft Matter \textbf{5}, 3106 (2009).
 \bibitem{Morrow}
 N. R. Morrow, J. Petrol. Technol. \textbf{42}, 1476 (1990).
 \bibitem{Liu}
 A. J. Liu, D. J. Durian, E. Herbolzheimer, and S. A. Safran,  Phys. Rev. Lett. 65, 1897–1900 (1990).
 \bibitem{Liu1}
 A. J. Liu, and G. S. Grest, Phys.Rev.A, \textbf{44}, 7894 (1991).
 \bibitem{Tanaka}
 H. Tanaka, Phys. Rev. Lett. \textbf{70}, 53 (1993).
 \bibitem{Iglauer}
 S. Iglauer, S. Favretto, G. Spinelli, G. Schena, and M. J. Blunt, Phys. Rev. E \textbf{82}, 056315 (2010).
 \bibitem{Pak}
 T. Pak, I. B. Butler, S. Geiger, M. I. J. van Dijke, and K. S. Sorbie, Proc. Natl. Acad. Sci. U.S.A. \textbf{112}, 1947 (2015).
 \bibitem{Strickland}
 B. Strickland, G. Leptoukh, and C. Roland, J. Phys. A: Math. Gen. \textbf{28}, L403 (1995). 
 \bibitem{Grunau}
 D. W. Grunau, T. Lookman, S. Y. Chen, and A. S. Lapedes, Phys. Rev. Lett. \textbf{71}, 4198 (1993).
 \bibitem{Chakrabarti}
 A. Chakrabarti, Phys. Rev. Lett. \textbf{69}, 1548 (1992).
\bibitem{Das}
S. K. Das, M. E. Fisher, J. V. Sengers, J. Horbach and K. Binder, Phys. Rev. Lett. \textbf{97}, 025702 (2006). 
\bibitem{Binder3}
K. Binder, in Phase Transformation of Materials, edited by R. W. Cahn, P. Haasen, and E. J. Kramer, Material Science and Technology (VCH, Weinheim, 1991), Vol. 5, p. 405; Kinetics of Phase Transitions, edited by S. Puri and V. Wadhawan (CRC Press, Boca Raton, 2009).
\bibitem{Porod}
G. Porod, in Small Angle X-Ray Scattering (ed. O. Glatter and L. Kratky, Academic Press, N. Y., 1983).
\bibitem{Gaurav}
G. P. Shrivastav, S. Krishnamoorthy, V. Banerjee and S. Puri, Europhys Lett. \textbf{96}, 36003 (2011).
\bibitem{Gaurav1}
G. P. Shrivastav, V. Banerjee and S. Puri, Eur. Phys. J. E (2014) Eur. Phys. J. E, 37: 98, 98 (2014).
\bibitem{Shaista}
S. Ahmad, S. Puri and S. K. Das, Phys. Rev. E \textbf{90}, 040302(R) (2014).
\bibitem{Lai}
Z. W. Lai, G. F. Mazenko and O. T. Valls, Phys. Rev. B \textbf{37}, 9481 (1988).
\bibitem{Corberi}
F. Corberi, E. Lippiello, A. Mukherjee, S. Puri and M. Zannetti, Phys. Rev. E \textbf{85}, 021141 (2012).
 \bibitem{Siggia}
E. D. Siggia, Phys. Rev. A \textbf{20}, 595 (1979).
\bibitem{Furukawa}
H. Furukawa, Phys. Rev. A \textbf{31}, 1103–1108 (1985).
\bibitem{skd}
S.K. Das, S. Roy, S. Majumdar, and S. Ahmad, Europhys. Lett. \textbf{97}, 66006 (2012).
\bibitem{Fisher1}
D.S. Fisher and D.A. Huse, Phys. Rev. B \textbf{38}, 373 (1988).
\bibitem{sahmed}
S. Ahmad, F. Corberi, S. K. Das, E. Lippiello, S. Puri, and M. Zannetti, Phys. Rev. E \textbf{86}, 061129 (2012).
 
\end{thebibliography}
\end{document}